\documentclass[aps,pre,twocolumn,showpacs]{revtex4}
\usepackage{epsfig}
\usepackage{times}
\begin{document}

\title{Phase transition and selection in a four-species cyclic
Lotka-Volterra model}
\author{Gy\"orgy Szab\'o}
\affiliation
{Research Institute for Technical Physics and Materials Science
P.O. Box 49, H-1525 Budapest, Hungary}
\author{Gustavo Arial Sznaider}
\affiliation
{Centro de Servicios Inform\'aticos and IFEVA,
Faculty of Agronomy, University of Buenos Aires,
Av. San Martin 4453, Buenos Aires (1417), Argentina}

\date{\today}

\begin{abstract}
We study a four species ecological system with cyclic dominance whose
individuals are distributed on a square lattice. Randomly chosen individuals
migrate to one of the neighboring sites if it is empty or invade this site
if occupied by their prey. The cyclic dominance maintains the coexistence of
all the four species if the concentration of vacant sites is lower than a
threshold value. Above the treshold, a symmetry breaking ordering occurs
via growing domains containing only two neutral species inside.
These two neutral species can protect each other from the external invaders
(predators) and extend their common territory. According to our Monte Carlo
simulations the observed phase transition is equivalent to those found in
spreading models with two equivalent absorbing states although the present
model has continuous sets of absorbing states with different portions of the
two neutral species. The selection mechanism yielding symmetric phases is
related to the domain growth process whith wide boundaries where the four
species coexist. 
\end{abstract}

\maketitle

Multispecies ecological models with spatial extension exhibit a large
variety of possible stationary states as well as phase transitions when
tuning the model parameters. In the original Lotka-Volterra models
\cite{lotka:pnas20,volterra:31} as well as in the generalized versions 
the spatial distribution of species is neglected 
(see \cite{hofbauer:98,drossel:ap01} for reviews). 
Now we report a phenomenon underlying the role of spatial effects
in the biological evolution.

In the simplest spatial Lotka-Volterra models 
the individuals of competitive species are residing on the sites of a 
lattice and the system evolution is governed by invasions along the nearest
neighbor links. In many cases the species form domains whith growing sizes
and sooner or later only one species will survive. Significantly different
behavior is found if the species dominate cyclically each other, i.e., the
corresponding food web is characterized by a directed ring graph
\cite{tainaka:prl89,tainaka:pre94,frachebourg:jpa98}.
Frachebourg and Krapivsky \cite{frachebourg:jpa98} have shown that fixation
occurs if the number of species $N_s$ exceeds a threshold value $N_f(d)$
depending on the spatial dimension $d$. In this case the species form
a frozen domain structure \cite{frachebourg:jpa98}. Conversely
[$N_s\leq N_f(d)$], the moving invasion fronts maintain a self-organizing
polydomain structure. These patterns are widely studied for $N_s=3$ 
\cite{tainaka:pre94,szabo:pre99,szabo:pre02a} because it can provide a
stability against some external invaders for the spatial models
\cite{boerlijst:pd91,szabo:pre01a,szabo:pre01b}. Sato {\it et al.}
\cite{sato:amc02} have
shown that, if only one of the invasion rates differs from unity for
even $N_s$, then only the species with odd (even) labels survive .
Very recently,
the species biodiversity were studied by similar models in bacterial
\cite{kerr:nature02}, phytoplankton \cite{huisman:e01} systems.

In the above lattice models each site is occupied by an individual of the
competitive species. Now we will consider a diluted version of these models
for $N_s=4$. Namely, the sites may be empty and the individuals are allowed
to jump to these empty sites. These elementary events can result in the
formation of ``defensive alliances'' consisting of two neutral species.
These two-species mixed states can preserve their territory from the
external invaders belonging to the remaining two species. Thus,
beside the above mentioned four-species state, this model has two sets of
``defensive alliances'' whose confrontations will determine the final
stationary state. When increasing the concentration of vacant sites this
system undergoes a phase transition from the symmetric four-species state
to one of the symmetric defensive alliances. This transition will be
interpreted by considering the average displacement (and velocity) of
boundary separating two competitive domains. 

In the present model the site $i$ of a square lattice can be empty ($s_i=0$)
or single occupied by one the four species (i.e., $s_i=1$, 2, 3, and 4)
dominating cyclically each other (1 beats 2 beats 3 beats 4 beats 1). The
time evolution is controlled by subsequent jumps or invasions at randomly
chosen nearest neighbor sites $i$ and $j$. The individual will jump to the
empty site, i.e., the value of $s_i$ and $s_j$ are exchanged ($s_i 
\leftrightarrow s_j$) if $s_i=0$ and $s_j > 0$ or $s_i >0$ and $s_j=0$.
Invasion occurs if predator and prey meet. For example, both the (1,2) and
(2,1) pairs transform into (1,1) [the further elementary invasions are given
by cyclic permutation of the species labels]. Nothing happens if $s_i=s_j$
as well as for neutral pairs, i.e., the pairs (1,3), (3,1), (2,4), and (4,2)
remain unchanged. The system is started from a random initial state. After
some transient time the system reaches a stationary state we study.

Notice that the above elemantary rules leave the number of vacant sites
unchanged and their distribution becomes uncorrelated after a suitable
relaxation time. The states containing only one species are considered
as absorbing states because the above rule does not create new species.
Besides this, the mixed states containing only two neutral species
(1+3 or 2+4) are also absorbing states and will be denoted as $D_{13}$
and $D_{24}$. In these stationary states the ratio of
the two species remains constant. In the presence of vacant sites the
migration eliminates the spatial correlations. For small sizes this
system can easily reach one of these absorbing states and afterwards
it stays there forever.

Our Monte Carlo (MC) simulations are performed on a square box with periodic
boundary conditions. In order to avoid the above mentioned small size
effect the linear size is varied from $L=400$ to 2000. The systematic 
simulations are started from a random initial state for different 
concentration of vacant sites ($\rho_0$). Within a time unit (MCS) each
pair has a chance once on the average to modify the state at one of the
corresponding sites. During the simulations we have
recorded the concentration of species and the pair configuration
probabilities on the nearest neighbor sites. Averaging over a suitable
sampling time interval we have determined the average species concentrations
($\rho_{\alpha}$, $\alpha = 1$, 2, 3, and 4). Furthermore, we have deduced
two quantities $P_{pp}$ and $P_n$ describing the probability of finding
predator-prey and neutral pairs on two nearest neighbor sites. Evidently,
$P_{pp}$ measures the invasion activity that vanishes in the absorbing
states.

\begin{figure}
\centerline{\epsfig{file=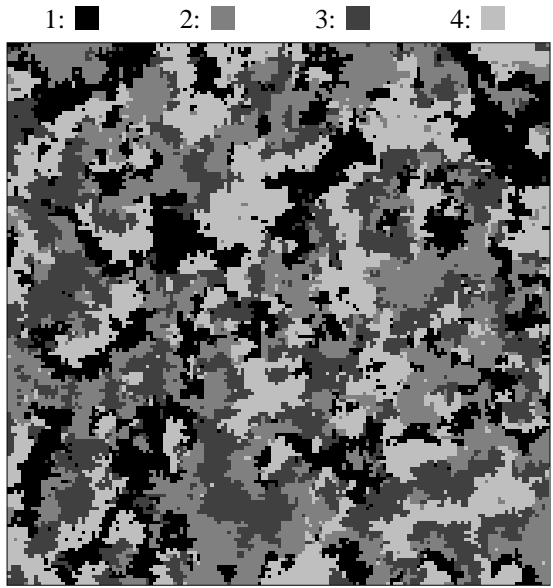,width=7.5cm}}
\caption{\label{fig:sp4s}Spatial distribution of the four species on the
square lattice if $\rho_0=0$. The grayscale of the four species are
indicated above the snaphot.}
\end{figure}

The visualization of species distribution shows a self-organizing polydomain
structure in the absence of vacant sites (for a typical snapshot see 
Fig.~\ref{fig:sp4s}). The species occur cyclically at each sites, however,
the short range interaction can not synchronize these accidental events.
In the pattern evolution one can easily recognize the traveling invasion
fronts that play crucial role in the maintenance of this polydomain 
structure \cite{tainaka:pre94,szabo:pre02a}. Similar spatiotemporal
patterns can be observed for low concentration of vacant sites.
Henceforth this spatiotemporal pattern is called $C$ state.

\begin{figure}
\centerline{\epsfig{file=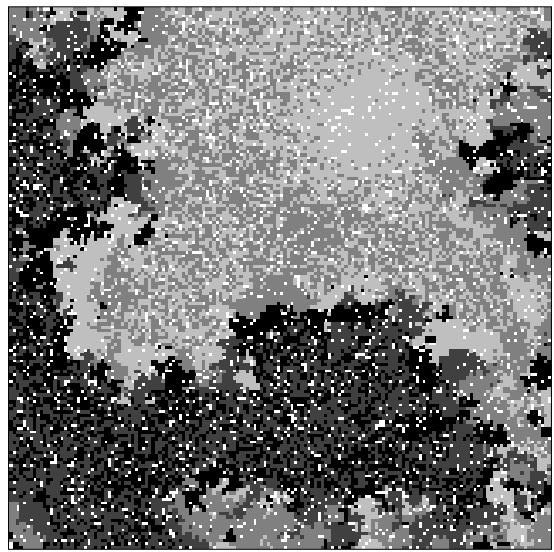,width=7.5cm}}
\caption{\label{fig:spdom}Typical domain structure at time $t=3000$ MCS
(Monte Carlo steps per sites) if initially ($t=0$) the spatial distribution
was random for $\rho_0=0.1$. The white boxes refer to empty sites while the
grayscale of species as in Fig.~\ref{fig:sp4s}.}
\end{figure}

In the stationary $C$ state $\rho_1=\rho_2=\rho_3=\rho_4=
(1-\rho_0)/4$ due to the cyclic symmetry. Strikingly different behavior 
occurs if $\rho_0 > \rho_{cr}=0.0623(1)$. When using lighter (darker)
grayscales for the species 1 and 3 (2 and 4) two types of growing domains
(namely $D_{13}$ and $D_{24}$) can be distinguished as shown in
Fig.~\ref{fig:spdom}. 
These growing domains are separated by wide regions of $C$ states. 
The growth process is similar to those observed in systems with two 
equivalent absorbing states 
\cite{dickman:pre95,hinrichsen:pre97,mehta:pre99,dornic:prl01}. Finally
the present system develops into one of the symmetric two-species absorbing
states $D_{13}$ or $D_{24}$ where $\rho_1=\rho_3=(1-\rho_0)/2$ and 
$\rho_2=\rho_4=0$, or $\rho_2=\rho_4=(1-\rho_0)/2$ and $\rho_1=\rho_3=0$.
The time of transition toward one of these states depends on $\rho_0$ and
$L$.
Both the species 1 and 3 benefit their spatially mixed coexistence because 
they protect each other from the external invasions. For example, 
the species 2 can invade the sites occupied originally by species 3,
however, the neighboring species 1 strikes back and eliminates the
invaders 2. At the same time, species 3 protects 1 against 4. This is the
reason why this association is called defensive alliance. Due to the cyclic
symmetry species 2 and 4 can form a similar defensive alliance.

The formation of defensive alliances was already observed in some other
multispecies Lotka-Volterra model where the cyclic invasion itself has
provided the protection mechanism \cite{szabo:pre01a,szabo:pre01b}.
In the present model, however, the protection is due to the mixing of
neutral species via the jumps to empty sites.
\begin{figure}
\centerline{\epsfig{file=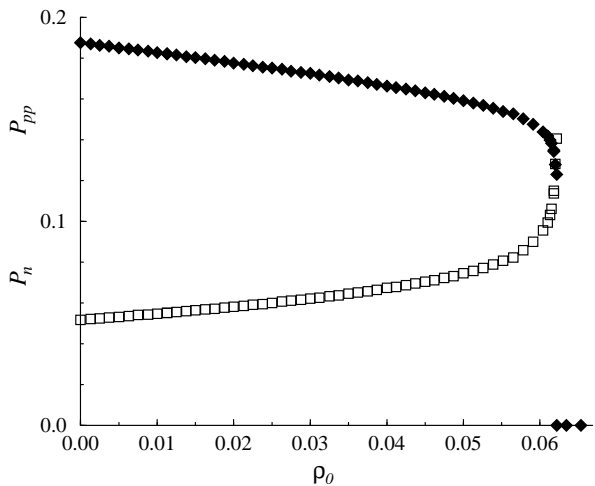,width=8cm}}
\caption{\label{fig:pairc0}Monte Carlo results for the probability of
finding predator-prey (closed diamonds) and neutral pairs (open squares)
on two nearest neighbor sites in the stationary states.}
\end{figure}

Figure \ref{fig:pairc0} demonstrates that the probability of neutral pairs
($P_n$) increases with the concentration of vacant sites in the stationary
states. Above the mentioned threshold value ($\rho_0 > \rho_{cr}$) this
quantity
tends to the uncorrelated value, $P_n=(1-\rho_0)^2/2$, characteristic to the
symmetric defensive alliance state. Simultaneously, the invasion activity
(or $P_{pp}$) decreases and drops suddenly to zero at $\rho_0=\rho_{cr}$.
The discontinuous transition is accompanied with enhanced fluctuations
(in all the quantities we studied) and a critical slowing down. To avoid
the undesired effects of fluctuation enhancements, the above MC data have
been obtained on large system ($L=2000$) with long relaxation and sampling
times ($t_r>10^4$ MCS and $t_s>10^5$ MCS) in the close vicinity of the
transition point.

The universal behavior of the nonequilibrium transitions into absorbing
states have extensively been studied for several decades (for a review see
the Refs. \cite{marro:99,hinrichsen:ap00}).
The dynamical systems with two equivalent absorbing states represent
a curious universality class (named after the voter model)
\cite{mendes:jpa94,dickman:pre95,munoz:pre97,hinrichsen:pre97,dornic:prl01}
whose general features are consistent to those described above. 

To have a deeper insight into the dynamics of the present model we now will 
study the displacement of interfaces separating the $C$ state and one
of the stationary defensive alliances ($D_{13}$ or $D_{24}$).
For the preparation of such an artificial domain structure the whole area
(torus) is divided into parallel strips with width of 500 lattice units.
The MC simulation is started from a random initial state (as above) for
$L=4000$ and after a suitable relaxation time $t_r$ uncorrelated $D_{13}$
state is created in every second strip. More precisely, species 1 and 3
are substituted randomly for the occupied sites located inside the 
corresponding strips. First we consider the results obtained for symmetric 
distribution, i.e., when inside the defensive alliances ($D_{13}$) 
$\rho_1=\rho_3=(1-\rho_0)/2$. The expansion (or shrinking) of the $C$ domains
can be monitored by evaluating the quantity
$\Phi(t) = \rho_1(t)-\rho_2(t)+\rho_3(t)-\rho_4(t)$. Notice that 
$\Phi$ vanishes ($\langle \Phi \rangle = 0$) for the
state $C$ whereas $\Phi(t)=\pm (1-\rho_0)$ in the absorbing states. 
The average displacement (measured in lattice unit) of the parallel
interfaces are derived straightforwardly from the variation of $\Phi(t)$.

\begin{figure}
\centerline{\epsfig{file=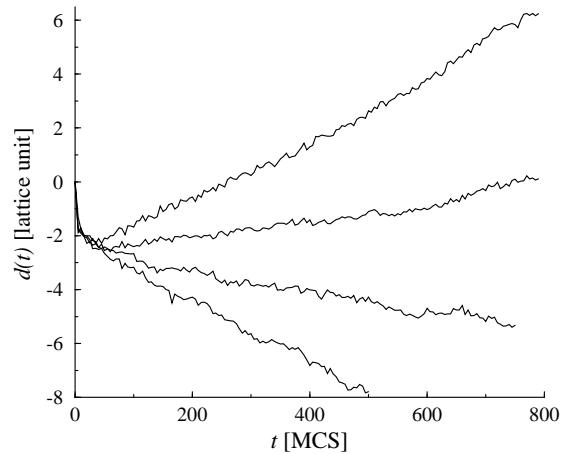,width=7.5cm}}
\caption{\label{fig:displ2}Average displacement of boundaries between the
states $C$ and $D_{13}$ as a function of time for $\rho_0=0.056$, 0.060, 
0.064, and 0.070 (from top to bottom).}
\end{figure}

Figure \ref{fig:displ2} displays the typical time dependences of the average
displacement $d(t)$ of the boundaries separating the $C$ and $D_{13}$ states.
The increase of $d(t)$ corresponds to the expansion of $C$ domains. The MC
data are obtained by averaging over 20 runs performed for $L=4000$ and
$t_r=3000$ MCS if $\rho_0 < \rho_{cr}$. Above the critical point
($\rho_0 > \rho_{cr}$) we have to use significantly shorter relaxation times
($t_r=200$ MCS) to avoid the difficulties caused by the appearance of the
$D_{13}$ and $D_{24}$ nucleons. It is remarkable that
at the beginning $d(t)$ decreases suddenly. After a suitable transient
time, however, the variation of $d(t)$ becomes linear and the fitted slope
can be interpreted as the average velocity $v$ of the invasion front.
 
\begin{figure}
\centerline{\epsfig{file=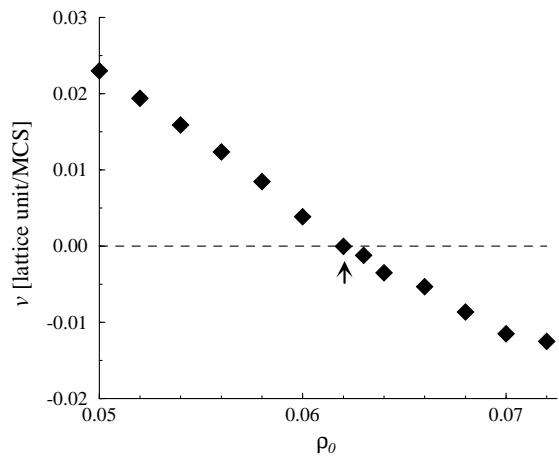,width=7.5cm}}
\caption{\label{fig:invvelm}Average velocity of invasion front between the
states $C$ and $D_{13}$ as a function of the concentration of vacant cites.
The arrow indicates the critical point derived from the investigation
of the stationary states. The statistical error is comparable to the
symbol size.}
\end{figure}

Figure \ref{fig:invvelm} clarifies that the $C$ state invades the
territories of defensive alliances  for low concentration of vacant sites.
The average invasion velocity decreases monotonously with $\rho_0$ and
becomes zero at $\rho_0 = \rho_{cr}$. In agreement with the expectation,
the area of the $C$ domains shrinks for $\rho_0 > \rho_{cr}$.

The above simulations were repeated by choosing asymmetric compositions
(e.g., $\rho_1 > \rho_3$) within the $D_{13}$ state. It is found, that
the asymmetry has influenced only the short time behavior. For example,
if $\rho_1 \gg \rho_3$ then the $C$ state can invade fast ($v \sim 1$)
those neighboring patches occupied by only the species 1 in the $D_{13}$
domains. Consequently, in this case one can observe a sudden increase
(instead of decrease as plotted in Fig.~\ref{fig:displ2}) in $d(t)$.
In the subsequent linear region, however the average velocity $v$ becomes
independent of $(\rho_1-\rho_3)$ within the statistical error. The
visualization of the species distribution has indicated that the boundary
between the $C$ and $D_{13}$ domains fluctuates very intensively. In 
fact, it is not a well defined boundary because the sites occupied by
species 1 and 3 can belong to both phases within a boundary layer.
Within this boundary layer the
cyclic invasions sustain the equivalence between $\rho_1$ and $\rho_3$
on average for long times. This boundary layer can be considered as a
symmetric species reservoir that drives an equalization between the
different species concentration in the asymmetric $D_{13}$ phase
via diffusion. 

This scenario is checked by considering the evolution from such an
initial state where the parallel strips (created as above) are filled
alternately by the symmetric $D_{24}$ and asymmetric $D_{13}$ states.
The simulations (for $\rho_0>\rho_{cr}$) have confirmed that the variation
of $d(t)$ is similar to a random walk meantime the difference
$\rho_1-\rho_3$ tends to zero for long times. This is the reason why we
have always found symmetric $D_{13}$ or $D_{24}$ states after the domain
coarsening process for sufficiently large system sizes.
At the same time this phenomenon can be interpreted as a selection
mechanism favorizing the symmetric defensive alliances. 

Here it is worth mentioning that the traditional mean-field and pair
approximations (for details see \cite{marro:99,satulovsky:pre94})
are capable to reproduce the existence of the above 
mentioned absorbing states. However, these techniques are not capable to
describe the observed phase transition. We think that this failure is due
to the very complex mechanisms consisting of many elementary steps
within a local cycle.

In summary, our work shows that a slight migration in the lattice
Lotka-Volterra models may significantly affect the species biodiversity.
Dilution and migration are attributes usually found in ecosystems
whose description should include these features.
The present model exemplifies that the migration of species
supports the formation defensive alliances in the multispecies ecological
systems. Furthermore, the confrontation between the different associations
of species plays crucial role in the selection of the survival population
structure. In this case a phase transition occurs when the rate of migration
is increased by allowing more and more vacant sites (and jumps) on the
lattice.

The above described features are observed for many other systems.
Preliminary results indicate clearly that similar behavior occurs if
the mixing is provided by the site exchange for neutral pairs without
introducing vacant sites. Furthermore, quantitatively similar behavior
is found for the continuous version of the present model, i.e., when
the individuals move freely on a planar surface and they create an offspring
if they eat a prey caught within a short distance. In fact, this former
finding inspired us to introduce a simpler model for the more rigorous
analysis.

We think that the mixing of neutral species can results in
other defensive alliances in multispecies systems. For example,
two equivalent alliances are expected to emerge in the $N_s$-species model
with a circular food web for even $N_s$. Evidently, such alliances can
occur for more complicated food webs when the species have several preys
and predators \cite{szabo:pre01a,szabo:pre01b}. In these situations the
competition between the possible (defensive) alliances will affect
the evolution of the ecological system including the food web itself
\cite{holt:er02,chowdhury:cm03}.

\begin{acknowledgments}
Thanks to Diego Gomez Deck with whom one of us developed the earlier
continuous version of the model. This work was supported by the Hungarian
National Research Fund under Grant No. T-33098.
\end{acknowledgments}


\begin{thebibliography}{26}
\expandafter\ifx\csname natexlab\endcsname\relax\def\natexlab#1{#1}\fi
\expandafter\ifx\csname bibnamefont\endcsname\relax
  \def\bibnamefont#1{#1}\fi
\expandafter\ifx\csname bibfnamefont\endcsname\relax
  \def\bibfnamefont#1{#1}\fi
\expandafter\ifx\csname citenamefont\endcsname\relax
  \def\citenamefont#1{#1}\fi
\expandafter\ifx\csname url\endcsname\relax
  \def\url#1{\texttt{#1}}\fi
\expandafter\ifx\csname urlprefix\endcsname\relax\def\urlprefix{URL }\fi
\providecommand{\bibinfo}[2]{#2}
\providecommand{\eprint}[2][]{\url{#2}}

\bibitem[{\citenamefont{Volterra}(1931)}]{volterra:31}
\bibinfo{author}{\bibfnamefont{V.}~\bibnamefont{Volterra}},
  \emph{\bibinfo{title}{Lecon sur la Theorie Mathematique de la Lutte pour la
  Vie}} (\bibinfo{publisher}{Gouthier-Villars}, \bibinfo{address}{Paris},
  \bibinfo{year}{1931}).

\bibitem[{\citenamefont{Lotka}(1920)}]{lotka:pnas20}
\bibinfo{author}{\bibfnamefont{A.~J.} \bibnamefont{Lotka}},
  \bibinfo{journal}{Proc. Natl. Acad. Sci. USA} \textbf{\bibinfo{volume}{6}},
  \bibinfo{pages}{410} (\bibinfo{year}{1920}).

\bibitem[{\citenamefont{Hofbauer and Sigmund}(1998)}]{hofbauer:98}
\bibinfo{author}{\bibfnamefont{J.}~\bibnamefont{Hofbauer}} \bibnamefont{and}
  \bibinfo{author}{\bibfnamefont{K.}~\bibnamefont{Sigmund}},
  \emph{\bibinfo{title}{Evolutionary Games and Population Dynamics}}
  (\bibinfo{publisher}{Cambridge University Press},
  \bibinfo{address}{Cambridge}, \bibinfo{year}{1998}).

\bibitem[{\citenamefont{Drossel}(2001)}]{drossel:ap01}
\bibinfo{author}{\bibfnamefont{B.}~\bibnamefont{Drossel}},
  \bibinfo{journal}{Adv. Phys.} \textbf{\bibinfo{volume}{50}},
  \bibinfo{pages}{209} (\bibinfo{year}{2001}).

\bibitem[{\citenamefont{Tainaka}(1994)}]{tainaka:pre94}
\bibinfo{author}{\bibfnamefont{K.}~\bibnamefont{Tainaka}},
  \bibinfo{journal}{Phys. Rev. E} \textbf{\bibinfo{volume}{50}},
  \bibinfo{pages}{3401} (\bibinfo{year}{1994}).

\bibitem[{\citenamefont{Frachebourg and Krapivsky}(1998)}]{frachebourg:jpa98}
\bibinfo{author}{\bibfnamefont{L.}~\bibnamefont{Frachebourg}} \bibnamefont{and}
  \bibinfo{author}{\bibfnamefont{P.~L.} \bibnamefont{Krapivsky}},
  \bibinfo{journal}{J. Phys. A: Math. Gen.} \textbf{\bibinfo{volume}{31}},
  \bibinfo{pages}{L287} (\bibinfo{year}{1998}).

\bibitem[{\citenamefont{Tainaka}(1989)}]{tainaka:prl89}
\bibinfo{author}{\bibfnamefont{K.}~\bibnamefont{Tainaka}},
  \bibinfo{journal}{Phys. Rev. Lett.} \textbf{\bibinfo{volume}{63}},
  \bibinfo{pages}{2688} (\bibinfo{year}{1989}).

\bibitem[{\citenamefont{Szab{\'o} et~al.}(1999)\citenamefont{Szab{\'o}, Santos,
  and Mendes}}]{szabo:pre99}
\bibinfo{author}{\bibfnamefont{G.}~\bibnamefont{Szab{\'o}}},
  \bibinfo{author}{\bibfnamefont{M.~A.} \bibnamefont{Santos}},
  \bibnamefont{and} \bibinfo{author}{\bibfnamefont{J.~F.~F.}
  \bibnamefont{Mendes}}, \bibinfo{journal}{Phys. Rev. E}
  \textbf{\bibinfo{volume}{62}}, \bibinfo{pages}{1095} (\bibinfo{year}{1999}).

\bibitem[{\citenamefont{Szab{\'o} and Szolnoki}(2002)}]{szabo:pre02a}
\bibinfo{author}{\bibfnamefont{G.}~\bibnamefont{Szab{\'o}}} \bibnamefont{and}
  \bibinfo{author}{\bibfnamefont{A.}~\bibnamefont{Szolnoki}},
  \bibinfo{journal}{Phys. Rev. E} \textbf{\bibinfo{volume}{65}},
  \bibinfo{pages}{036115} (\bibinfo{year}{2002}).

\bibitem[{\citenamefont{Boerlijst and Hogeweg}(1991)}]{boerlijst:pd91}
\bibinfo{author}{\bibfnamefont{M.~C.} \bibnamefont{Boerlijst}}
  \bibnamefont{and} \bibinfo{author}{\bibfnamefont{P.}~\bibnamefont{Hogeweg}},
  \bibinfo{journal}{Physica D} \textbf{\bibinfo{volume}{48}},
  \bibinfo{pages}{17} (\bibinfo{year}{1991}).

\bibitem[{\citenamefont{Szab{\'o} and
  Cz{\'a}r{\'a}n}(2001{\natexlab{a}})}]{szabo:pre01a}
\bibinfo{author}{\bibfnamefont{G.}~\bibnamefont{Szab{\'o}}} \bibnamefont{and}
  \bibinfo{author}{\bibfnamefont{T.}~\bibnamefont{Cz{\'a}r{\'a}n}},
  \bibinfo{journal}{Phys. Rev. E} \textbf{\bibinfo{volume}{63}},
  \bibinfo{pages}{061904} (\bibinfo{year}{2001}{\natexlab{a}}).

\bibitem[{\citenamefont{Szab{\'o} and
  Cz{\'a}r{\'a}n}(2001{\natexlab{b}})}]{szabo:pre01b}
\bibinfo{author}{\bibfnamefont{G.}~\bibnamefont{Szab{\'o}}} \bibnamefont{and}
  \bibinfo{author}{\bibfnamefont{T.}~\bibnamefont{Cz{\'a}r{\'a}n}},
  \bibinfo{journal}{Phys. Rev. E} \textbf{\bibinfo{volume}{64}},
  \bibinfo{pages}{042902} (\bibinfo{year}{2001}{\natexlab{b}}).

\bibitem[{\citenamefont{Sato et~al.}(2002)\citenamefont{Sato, Yoshida, and
  Konno}}]{sato:amc02}
\bibinfo{author}{\bibfnamefont{K.}~\bibnamefont{Sato}},
  \bibinfo{author}{\bibfnamefont{N.}~\bibnamefont{Yoshida}}, \bibnamefont{and}
  \bibinfo{author}{\bibfnamefont{N.}~\bibnamefont{Konno}},
  \bibinfo{journal}{Appl. Math. Comp.} \textbf{\bibinfo{volume}{126}},
  \bibinfo{pages}{255} (\bibinfo{year}{2002}).

\bibitem[{\citenamefont{Kerr et~al.}(2002)\citenamefont{Kerr, Riley, Feldman,
  and Bohannan}}]{kerr:nature02}
\bibinfo{author}{\bibfnamefont{B.}~\bibnamefont{Kerr}},
  \bibinfo{author}{\bibfnamefont{M.~A.} \bibnamefont{Riley}},
  \bibinfo{author}{\bibfnamefont{M.~W.} \bibnamefont{Feldman}},
  \bibnamefont{and} \bibinfo{author}{\bibfnamefont{B.~J.~M.}
  \bibnamefont{Bohannan}}, \bibinfo{journal}{Nature}
  \textbf{\bibinfo{volume}{418}}, \bibinfo{pages}{171} (\bibinfo{year}{2002}).

\bibitem[{\citenamefont{Huisman and Weissing}(2001)}]{huisman:e01}
\bibinfo{author}{\bibfnamefont{J.}~\bibnamefont{Huisman}} \bibnamefont{and}
  \bibinfo{author}{\bibfnamefont{F.~J.} \bibnamefont{Weissing}},
  \bibinfo{journal}{Ecology} \textbf{\bibinfo{volume}{82}},
  \bibinfo{pages}{2682} (\bibinfo{year}{2001}).

\bibitem[{\citenamefont{Dickman and Tretyakov}(1995)}]{dickman:pre95}
\bibinfo{author}{\bibfnamefont{R.}~\bibnamefont{Dickman}} \bibnamefont{and}
  \bibinfo{author}{\bibfnamefont{A.~Y.} \bibnamefont{Tretyakov}},
  \bibinfo{journal}{Phys. Rev. E} \textbf{\bibinfo{volume}{52}},
  \bibinfo{pages}{3218} (\bibinfo{year}{1995}).

\bibitem[{\citenamefont{Hinrichsen}(1997)}]{hinrichsen:pre97}
\bibinfo{author}{\bibfnamefont{H.}~\bibnamefont{Hinrichsen}},
  \bibinfo{journal}{Phys. Rev. E} \textbf{\bibinfo{volume}{55}},
  \bibinfo{pages}{219} (\bibinfo{year}{1997}).

\bibitem[{\citenamefont{Dornic et~al.}(2001)\citenamefont{Dornic, Chat{\'e},
  Chave, and Hinrichsen}}]{dornic:prl01}
\bibinfo{author}{\bibfnamefont{I.}~\bibnamefont{Dornic}},
  \bibinfo{author}{\bibfnamefont{H.}~\bibnamefont{Chat{\'e}}},
  \bibinfo{author}{\bibfnamefont{J.}~\bibnamefont{Chave}}, \bibnamefont{and}
  \bibinfo{author}{\bibfnamefont{H.}~\bibnamefont{Hinrichsen}},
  \bibinfo{journal}{Phys. Rev. Lett.} \textbf{\bibinfo{volume}{87}},
  \bibinfo{pages}{045701} (\bibinfo{year}{2001}).

\bibitem[{\citenamefont{Mehta and Luck}(1999)}]{mehta:pre99}
\bibinfo{author}{\bibfnamefont{A.}~\bibnamefont{Mehta}} \bibnamefont{and}
  \bibinfo{author}{\bibfnamefont{J.~M.} \bibnamefont{Luck}},
  \bibinfo{journal}{Phys. Rev. E} \textbf{\bibinfo{volume}{60}},
  \bibinfo{pages}{5218} (\bibinfo{year}{1999}).

\bibitem[{\citenamefont{Marro and Dickman}(1999)}]{marro:99}
\bibinfo{author}{\bibfnamefont{J.}~\bibnamefont{Marro}} \bibnamefont{and}
  \bibinfo{author}{\bibfnamefont{R.}~\bibnamefont{Dickman}},
  \emph{\bibinfo{title}{Nonequilibrium Phase Transitions in Lattice Models}}
  (\bibinfo{publisher}{Cambridge University Press},
  \bibinfo{address}{Cambridge}, \bibinfo{year}{1999}).

\bibitem[{\citenamefont{Hinrichsen}(2000)}]{hinrichsen:ap00}
\bibinfo{author}{\bibfnamefont{H.}~\bibnamefont{Hinrichsen}},
  \bibinfo{journal}{Adv. Phys.} \textbf{\bibinfo{volume}{49}},
  \bibinfo{pages}{815} (\bibinfo{year}{2000}).

\bibitem[{\citenamefont{Mendes et~al.}(1994)\citenamefont{Mendes, Dickman,
  Henkel, and Marques}}]{mendes:jpa94}
\bibinfo{author}{\bibfnamefont{J.~F.~F.} \bibnamefont{Mendes}},
  \bibinfo{author}{\bibfnamefont{R.}~\bibnamefont{Dickman}},
  \bibinfo{author}{\bibfnamefont{M.}~\bibnamefont{Henkel}}, \bibnamefont{and}
  \bibinfo{author}{\bibfnamefont{M.~C.} \bibnamefont{Marques}},
  \bibinfo{journal}{J. Phys. A: Math. Gen.} \textbf{\bibinfo{volume}{27}},
  \bibinfo{pages}{3019} (\bibinfo{year}{1994}).

\bibitem[{\citenamefont{Mu{\~n}oz et~al.}(1997)\citenamefont{Mu{\~n}oz,
  Grinstein, and Tu}}]{munoz:pre97}
\bibinfo{author}{\bibfnamefont{M.~A.} \bibnamefont{Mu{\~n}oz}},
  \bibinfo{author}{\bibfnamefont{G.}~\bibnamefont{Grinstein}},
  \bibnamefont{and} \bibinfo{author}{\bibfnamefont{Y.}~\bibnamefont{Tu}},
  \bibinfo{journal}{Phys. Rev. E} \textbf{\bibinfo{volume}{56}},
  \bibinfo{pages}{5101} (\bibinfo{year}{1997}).

\bibitem[{\citenamefont{Satulovsky and Tom{\'e}}(1994)}]{satulovsky:pre94}
\bibinfo{author}{\bibfnamefont{J.~E.} \bibnamefont{Satulovsky}}
  \bibnamefont{and} \bibinfo{author}{\bibfnamefont{T.}~\bibnamefont{Tom{\'e}}},
  \bibinfo{journal}{Phys. Rev. E} \textbf{\bibinfo{volume}{49}},
  \bibinfo{pages}{5073} (\bibinfo{year}{1994}).

\bibitem[{\citenamefont{Holt}(2002)}]{holt:er02}
\bibinfo{author}{\bibfnamefont{R.~D.} \bibnamefont{Holt}},
  \bibinfo{journal}{Ecol. Res.} \textbf{\bibinfo{volume}{17}},
  \bibinfo{pages}{261} (\bibinfo{year}{2002}).

\bibitem[{\citenamefont{Chowdhury and Stauffer}()}]{chowdhury:cm03}
\bibinfo{author}{\bibfnamefont{D.}~\bibnamefont{Chowdhury}} \bibnamefont{and}
  \bibinfo{author}{\bibfnamefont{D.}~\bibnamefont{Stauffer}},
  \eprint{cond-mat/0305322}.

\end{thebibliography}

\end{document}